\documentclass[12pt]{article}
\usepackage[hypertex]{hyperref}
\usepackage[dvipdfmx]{graphicx} 
\usepackage{amsmath,amssymb,amsfonts,bm,color,cite}

\def\d{{\bm d}}   \def\A{{\bm A}} \def\F{{\bm F}} 

\newcommand{\we}{\wedge}
\newcommand{\To}{\Rightarrow}

\newcommand{\getsto}{\leftrightarrow}

\newcommand{\tr}{{\rm tr}}
\newcommand{\vev}[1]{ \langle {#1} \rangle }

\newcommand{\lsp}{ \left ( }
\newcommand{\rsp}{ \right ) }
\renewcommand{\L}{\mathcal{L}}

\begin{document}

\begin{titlepage}

\begin{flushright}
KEK-TH-1791
\end{flushright}

\vskip 1.35cm

\begin{center}

{\large \bf A symmetry-breaking mechanism by parity assignment \\ in the noncommutative Higgs model}

\vskip 1.2cm

Masaki J.S. Yang

\vskip 0.4cm

{\it Institute of Particle and Nuclear Studies,\\
High Energy Accelerator Research Organization (KEK), \\
Tsukuba 305-0801, Japan\\
}

\date{\today}

%%%%%%%%%%%%%%%%%%%%%%%%%%%%%%%%%%%%%%%%%%%%%%%%
\begin{abstract} %%%%%%%%%%%%%%%%%%%%%%%%%%%%%%%%%%%%%%%
%%%%%%%%%%%%%%%%%%%%%%%%%%%%%%%%%%%%%%%%%%%%%%%%
We apply the orbifold grand unified theory (GUT) mechanism to the noncommutative Higgs model. 
An assignment of $Z_{2}$ parity to the  ``constituent fields'' induces 
parity assignments of both the gauge and Higgs bosons, 
because these bosons are treated as some kind of composite field in this formalism.
As a result, some of the gauge bosons and the colored triplet Higgs boson receive heavy mass comparable to the  GUT scale, and the gauge symmetry is broken. 
 No particles appear other than the SM ones in the massless states.

%%%%%%%%%%%%%%%%%%%%%%%%%%%%%%%%%%%%%%%%%%%%%%%
\end{abstract} %%%%%%%%%%%%%%%%%%%%%%%%%%%%%%%%%%%%%%%%
%%%%%%%%%%%%%%%%%%%%%%%%%%%%%%%%%%%%%%%%%%%%%%%

\end{center}
\end{titlepage}

%%%%%%%%%%%
\section{Introduction}
%%%%%%%%%%%

The discovery of the Higgs boson  \cite{Aad:2012tfa,Chatrchyan:2012ufa}, a great triumph of science, has completed the particle contents of the Standard Model (SM).
However, the theoretical origin of the Higgs boson is still unclear.
Innumerable theories and models have been considered so far, in order to explain the existence of the boson and clarify the underlying physics beyond the SM.

Among these attempts, one interesting possibility is the noncommutative geometry inspired by the Yang--Mills--Higgs model, developed by A. Connes \cite{Connes:1990qp, Connes:1994yd}. We call this model noncommutative Higgs model (NHM) for short.
In this picture, the underlying spacetime is assumed to be a multi-sheet of ordinary Minkowski space $M^{4} \times Z_{n}$. The Higgs boson is interpreted as a gauge boson along a discrete extra dimension, 
which has noncommutative differential algebra. 
For example, in the simplest model $M^{4} \times Z_{2}$, 
the coordinates are represented as $(x^{\mu}, y = \pm1)$. Thus, $y^{2} = 1$ holds, and it leads to the anti-commutative relation $ y \, dy = - dy \, y $. This noncommutative differential algebra generates  nonzero Higgs mass and the spontaneous symmetry breaking (SSB) mechanism.

This concept is widely followed by its successors \cite{DuboisViolette:1988ir, Coquereaux:1990ev, Sitarz:1993zf, Morita:1993zv} and explained in several reviews 
\cite{Dimakis:1993bq,Chamseddine:1994hm,ORaifeartaigh:1998pk,Castellani:2000xt}.
It is also applied to supersymmetry \cite{Chamseddine:1994np}, and
 the grand unified theories (GUT), e.g., 
SU(5) \cite{Chamseddine:1992kv, Chamseddine:1992nx, Morita:1993xj, Sogami:1996xy} 
or SO(10) models \cite{Chamseddine:1993is, Okumura:1995ih}.
Meanwhile, when this kind of model is regarded as a theory with an extra dimension, 
several mechanisms and structures in the normal extra dimension can be recast to the noncommutative extra dimension.

 As a typical example, in this paper, we implement the orbifold GUT mechanism by Y. Kawamura \cite{Kawamura:2000ev} in the SU(5) NHM.
In the previous standard studies of the NHM, the symmetry-breaking pattern is determined by selecting the {\it distance} matrices $M_{nm}$. 
By contrast, we imposed an assignment of $Z_{2}$ parity to the ``constituent fields''.
This induces parity assignments of both gauge and Higgs bosons, 
because these bosons are treated as some kind of composite field in this formalism.
This is the main difference from the original orbifold GUT model, which treats the assignments of bosons as independent conditions. 
As a result, some of the gauge bosons and the colored triplet Higgs boson receive heavy mass comparable to the  GUT scale, and the SU(5) gauge symmetry is broken. 
 No particles appear other than the SM ones in the massless states. 

This paper is organized as follows. 
In the next section, we review the generalized gauge theory in the NHM.
In Sect. 3, we present the symmetry-breaking mechanism by parity assignment, and discuss a new field and proton decay. Section 4 is devoted to the conclusion.

%%%%%%%%%%%%%%%%%%%%%%%%%
\section{Generalized gauge theory on $M^{4} \times Z_{n}$}
%%%%%%%%%%%%%%%%%%%%%%%%%

Here, we define several definitions and the main frame of the formulation in the generalized gauge theory on $M^{4} \times Z_{n}$. 
Since the formulation is the same as those in Ref. \cite{Morita:1993xj}, a detailed description is omitted.
The essential difference in the formulations between the original paper \cite{Chamseddine:1992kv, Chamseddine:1992nx} and Ref. \cite{Morita:1993xj} is the flavor dependence of the {\it distance} matrices $M_{nm}$; the differential algebraic structure is almost equivalent.

%%%%%%%%%%%%%%%%%%%%%%%%%%%%%%
\subsection{Differential calculus}
%%%%%%%%%%%%%%%%%%%%%%%%%%%%%%

On the $n$-sheeted Minkowski space $M^{4} \times Z_{n}$, 
the coordinates are represented by $(x^{\mu}, y = 1-N),$
and a function on the $n$th sheet is expressed as $f(x,n) \equiv f_{n}$.
In this space, the exterior derivative is generalized to $\d = d + d_{\chi}$, introducing extra exterior derivatives $d_{\chi}$ and the differential forms $\chi_{m}$. They are defined as follows:
\begin{align*}
\d f_{n} & =(d + d_{\chi}) f_{n}, ~~~~
 d \hspace{0.5pt} f_{n}  = \partial_\mu f_{n} dx^\mu, \\
 d_{\chi} f &= \sum_{m \neq n} d_{\chi_{m} } f = \sum_{m \neq n} [ M_{nm} f_{m}  - f_{n} M_{nm}] \chi_{m},  
\end{align*}
where the matrices $M_{nm}^{\dagger} = M_{mn} (n \neq m)$ represent the {\it distances} between two sheets.
In order to keep the nilpotency of the generalized exterior derivative $\d^{2} = 0$, the differential forms $dx^{\mu}$ and $\chi_{m}$ should be anti-commutative, $dx^{\mu} \we \chi_{m} = - \chi_{n} \we dx^{\mu}$ ,
and the $d_{\chi}$ should also be nilpotent, $d_{\chi}^{2} = 0.$ 
To prove $d_{\chi}^{2} = 0$, and to keep the differential algebra consistent, there are several ``index shifting rules'' between $f_{n}, M_{nm},$ and $\chi_{m}$. These shifting rules are the source of the noncommutativity that corresponds to the relation $y \, dy = - dy \, y $ in other formulations \cite{DuboisViolette:1988ir, Coquereaux:1990ev, Sitarz:1993zf, Morita:1993zv}.
The precise form of these rules and a proof of $d_{\chi}^{2} = 0$ is presented in Ref. \cite{Morita:1993xj}.

%%%%%%%%%%%%%%%%%%%%%%%%%%%%%%%%
\subsection{Generalized connection }
%%%%%%%%%%%%%%%%%%%%%%%%%%%%%%%%

In several classes of models \cite{DuboisViolette:1988ir, Coquereaux:1990ev, Sitarz:1993zf, Morita:1993zv}, the gauge and Higgs bosons are regarded as elemental fields. By contrast, Connes's original paper and its successors \cite{Connes:1990qp, Chamseddine:1992kv, Chamseddine:1992nx}  treat these bosons as some kind of composite field. This formulation is effective in constructing realistic GUT models in particular. 
Here we adopt this composite formulation, which is also quoted from Ref. \cite{Morita:1993xj}.
In this picture, a gauge field consists of ``constituent fields'', defined on the same sheet, while a Higgs field is defined on the different sheets. 
The quotation marks ``~'' mean that the detailed dynamics of the binding mechanism is not specified.

The above picture is described by utilizing the generalized connection one-form $\A$, defined as
\begin{align}
\A (x,n) &\equiv \sum_{i} a_{i}^{\dagger} (x,n) \d a_{i} (x,n) , \nonumber \\
&= \sum_{i} a_{i}^{\dagger} (x,n) da_{i}(x,n) + \sum_{i} \sum_{m \neq n} a_{i}^{\dagger} (x,n) 
[M_{nm} a_{i} (x,m) - a_{i} (x,n) M_{nm}] \chi_{m} \nonumber , \\
& \equiv A(x,n) + \sum_{m\neq n} \Phi_{nm} (x) \chi_{m} .
\label{ga}
\end{align}
Here, the ``constituent field'' $a_{i} (x,n)$ is a square-matrix-valued continuous function defined on the $n$th sheet, and the summation over $i$ is assumed to be a finite sum.
The last line in Eq.~(\ref{ga}) defines the gauge and Higgs fields as follows:
\begin{align}
A(x,n) &= \sum_{i} a_{i}^{\dagger} da_{i} (x,n) , \\
\Phi_{nm} (x) &= \sum_{i} a_{i}^{\dagger} (x,n) [M_{nm} a_{i} (x,m) - a_{i} (x,n) M_{nm}] ~~~ (n\neq m).
\label{Phi}
\end{align}
Henceforth, we use the notation $\A(x,n) = \A_{n}, A(x,n) = A_{n},$ and $a_{i} (x,n) = a^{i}_{n}$ as shortened forms.
As in Refs. \cite{Chamseddine:1992kv, Chamseddine:1992nx, Morita:1993xj},  we impose the following Hermitian condition
\begin{align}
& \sum_{i} a^{i \, \dagger}_{n} a^{i}_{n} = 1 , ~~  \To ~~
A_{n}^{\dagger} = -A_{n}, ~~~ (\sum_{m \neq n}\Phi_{nm} \chi_{m})^{\dagger} = - \sum_{m \neq n} \Phi_{mn} \chi_{n} .
\label{Hermitian}
\end{align}
From Eq.~(\ref{Hermitian}), we assume that $\Phi_{nm}^{\dagger} = \Phi_{mn}$ and $\chi_{m}^{\dagger} = - \chi_{m}$.
For later convenience, the {\it back-shifted} Higgs field 
\begin{equation}
H_{nm} \equiv \Phi_{nm} + M_{nm} = \sum_{i} a^{i \, \dagger}_{n} M_{nm} a^{i}_{m} ,
\label{backshift}
\end{equation}
is also introduced.

The field-strength two-form is defined as follows:
\begin{equation}
\F_{n} = \d \A_{n} + \A_{n} \we \A_{n}  ,
\label{Fn}
\end{equation}
where
\begin{equation}
\d \A_{n} = \sum_{i} \d a^{i \,\dagger}_{n} \we \d a^{i}_{n} ,
\end{equation}
because of $\d^{2} = 0$. 
After some calculation, the explicit form of $\F_{n}$ contains the following three elements:
\begin{align}
\F_{n} &= F_{n \, \mu\nu} dx^{\mu} \we dx^{\nu} + \sum_{m \neq n} D_{\mu} H_{nm} \, dx^{\mu} \we \chi_{m}  
+ \sum_{m\neq n , l \neq m} (H_{nm} H_{ml} - X'_{nml}) \chi_{m} \we \chi_{l} ,
\label{threeF}
\end{align}
where
\begin{align}
F_{\mu\nu} (x,n) \equiv F_{n \, \mu\nu} &= {1\over 2} \lsp \partial_{\mu} A_{\nu} (x,n) - \partial_{\nu} A_{\mu} (x,n) + [A_{\mu} (x,n), A_{\mu} (x,n)  ] \rsp \\
D_{\mu} H_{nm} (x) &= \partial_{\mu} H_{nm} (x) + A_{\mu} (x,n) H_{nm} (x) - H_{nm}   (x) A_{\mu} (x,m) , \\
X_{nml}' (x) &= \sum_{i} a^{\dagger}_{i} (x,n) M_{nm} M_{ml} a_{i}  (x,l).
\label{Xnml}
\end{align}
It contains the gauge boson, the Higgs boson, and the new field $X_{nml}' (x)$.
The treatment of $X'_{nml}$ is decided by whether $X'_{nml}$ is a dependent function of the Higgs fields $H_{nm}$ or not.
If $X'_{nml}$ can be written as some function of the Higgs fields $X'_{nml} = f (H_{nm})$, $X'_{nml}$ is treated as a Higgs interaction terms. If not, $X'_{nml}$ is regarded as an {\it auxiliary field} that does not have a kinetic term, and then it will be eliminated from the Lagrangian by the equation of motion $\partial \L / \partial X_{nml}' = 0$.  

In order to determine the gauge transformation of ``composite fields'',  
the ``constituent field'' $a^{i}_{n}$ is first assigned to be a fundamental representation under the $n$th  gauge transformation:
\begin{equation}
a^{i \,'}_{n} = a^{i}_{n} g_{n} ,
\label{transai}
\end{equation}
where $g_{n} = g(x,n) = (g (x,n)^{\dagger})^{-1}$ is an arbitrary unitary matrix associated with the gauge group on the sheet $n$. 
From Eq.~(\ref{transai}),  the generalized connection and field strength will transform as the standard form:
\begin{align}
\A^{'}_{n} &= g^{-1}_{n} \A_n g_{n} + g^{-1}_{n} \d g_{n} , \label{transA} \\
\F^{'}_{n} &= g^{-1}_{n} \F_{n} \, g_{n} , 
\end{align}
with $M_{nm}^{'} = M_{nm}$. In addition, 
Eq.~(\ref{transA}) implies that the back-shifted Higgs field (\ref{backshift}) transforms as a bifundamental representation:
\begin{equation}
H_{nm}^{'} = g^{-1}_{n} H_{nm} \, g_{m} .
\end{equation}
%

%%%%%%%%%%%%%%%%%
\subsection{The Lagrangian}
%%%%%%%%%%%%%%%%%

In consequence, the gauge-invariant Lagrangian is formulated by 
\begin{equation}
\L_{\rm YMH} = - {1\over 4} \sum_{n} {1\over g_{n}^{2}} \tr \vev{\F_{n} , \F_{n}} ,
\label{LYMH}
\end{equation}
where independent coupling constants $g_{n}$ are introduced for gauge fields on each $n$th sheet. 

In order to calculate Eq.~(\ref{LYMH}), 
the metric of the space $M^{4} \times Z_{n}$ is specified as
\begin{align*}
&\vev{dx^{\mu}, dx^{\nu} } = g^{\mu\nu},  \\
& \vev{\chi_{n}, dx^{\mu}} = 0 , \\
&\vev{\chi_{n}, \chi_{m}} = - \delta_{nm} \alpha_{n}^{2} ,
\end{align*}
where $g^{\mu\nu} = {\rm diag} (+,-,-,-)$. 
Then the inner products of the two-forms are found to be
\begin{align}
\vev{dx^\mu \wedge dx^\nu, dx^\rho \wedge dx^\sigma} &=g^{\mu\rho}g^{\nu\sigma}- g^{\mu\sigma}g^{\nu\rho},\\
\vev{dx^\mu \wedge \chi_{m}, dx^\nu \wedge \chi_{n}} &=- \delta_{nm} \alpha_{n}^2 g^{\mu\nu}  ,\\
\vev{\chi_{k} \wedge \chi_{l},\chi_{n} \wedge \chi_{m}}&= \alpha_{n}^2 \alpha_{m}^{2} (\delta_{kn} \delta_{lm} - \delta_{km} \delta_{ln}) ,
\end{align}
while the other inner products among the basis two-forms are found to be zero.
Finally, we split $X_{nml}(x)$ in Eq.~(\ref{Xnml}) into two terms according to $n = l$ and $n \neq l$ for convenience:
\begin{align}
P_{nm} (x) &= H_{nm} (x) H_{mn} (x) - X'_{nmn} (x), \\
Q_{nml} (x) &= H_{nm} (x) H_{ml} (x) - X'_{nml} (x), 
\label{26}
\end{align}
where $n \neq m \neq l \neq n$. 
From Eq.~(\ref{LYMH}), we obtain the final expression of the (bosonic sector of the) Lagrangian \cite{Morita:1993xj}:
\begin{align}
\L_{\rm YMH} = &- \sum_{n} { 1\over 8 g_{n}^{2}} \tr | F_{\mu\nu} (x,n) |^{2} 
+ \sum_{n} \sum_{n \neq m} {\alpha_{m}^{2} \over 4 g_{n}^{2}} \tr |D_{\mu} H_{nm}(x)|^{2} \nonumber \\
&- \sum_{n} \sum_{n \neq m} { \alpha_{m}^{2} \alpha_{n}^{2} \over 4 g_{n}^{2}} \tr |P_{nm}(x)|^{2} 
- \sum_{n} \sum_{ m\neq n , l \neq n} { \alpha_{m}^{2} \alpha_{l}^{2} \over 4 g_{n}^{2}} \tr |Q_{nml}(x)|^{2} .
\label{finalL}
\end{align}
The Lagrangian (\ref{finalL}) is subdivided into four terms corresponding to the decomposition~(\ref{threeF}): 
The first term is the pure Yang--Mills term with independent coupling constants, 
the second is the Higgs kinetic energy term, 
the third represents the self-coupling of Higgs $H_{nm}$, 
and the fourth term describes interactions among different Higgs $H_{nm}$ and $H_{ml}$. 

%%%%%%%%%%%%%%%%%%%%%%%%%%%%%%
\section{SU(5) GUT-breaking by parity assignment}
%%%%%%%%%%%%%%%%%%%%%%%%%%%%%%

In this section, we review an SU(5) GUT in the NHM briefly, and implement the orbifold GUT mechanism in this toy model.
 In the formalism presented in the previous section, an SU(5) GUT model requires $N \geqq 3$.
This is because $N \geqq 3$ realizes more than two independent $M_{nm}$, which correspond to  two energy scales of SU(5) symmetry breaking, SU(5) $\to$ SU(3)$_{c}$ $\times$ SU(2)$_{L}$ $\times$ U(1)$_{Y}$ and SU(3)$_{c}$ $\times$ SU(2)$_{L}$ $\times$ U(1)$_{Y}$ $\to$ SU(3)$_{c}$ $\times$ U(1)$_{\rm em}$. 
Then we choose $N=3$ to describe the SU(5) GUT \cite{Chamseddine:1992kv, Morita:1993xj}. The indices $n, m, l$ in Eq.~(\ref{finalL}) then take the values 1, 2, 3 only. 

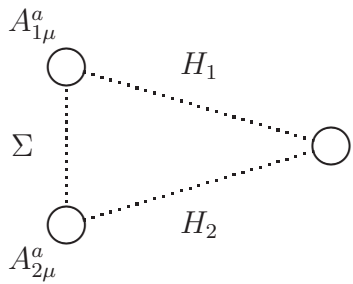
\begin{figure}[h!]
\begin{center}  \begin{picture}(260,110)
\thicklines  
\put(80,80){\circle{14}}  
\put(80,20){\circle{14}}   \put(180,50){\circle{14}}
\qbezier[20](80,20)(80,50)(80,80)
\qbezier[30](80,80)(130,65)(180,50)
\qbezier[30](80,20)(130,35)(180,50)
\put(80,80){\color{white}{\circle*{13}}}   
\put(80,20){\color{white}{\circle*{13}}}   \put(180,50){\color{white}{\circle*{13}}}  
\put(63,50){\makebox(0.4,0.6){$\Sigma$}}
\put(67,95){\makebox(0.4,0.6){$A_{1 \mu}^{a}$}}   \put(130,80){\makebox(0.4,0.6){$H_{1}$}}
\put(67,5){\makebox(0.4,0.6){$A_{2 \mu}^{a}$}}   \put(130,20){\makebox(0.4,0.6){$H_{2}$}}
\end{picture}   
\caption{A schematic image of the SU(5) model on $M^{4} \times Z_{3}$. }
\end{center}
\end{figure}

At first, the ``constituent fields'' $a^{i}_{1}, a^{i}_{2}$ are assumed to be complex 5 $\times$ 5 matrices and $a^{i}_{3}$ has a real-valued continuous function that satisfy Eq.~(\ref{Hermitian}), $\sum_{i} a^{i \, \dagger}_{n} a^{i}_{n} = 1$. 
Additionally, we consider a parity symmetry between $n=1$ and $n=2$ sheets of spacetime, 
and impose the following parity condition for the ``constituent fields'':
\begin{equation}
 a^{i}_{2} = P a^{i}_{1} P,
 \label{PP}
\end{equation}
where $P$ = diag $(-1,-1,-1,+1,+1)$.
In order to break the gauge symmetry, this parity assignment is found to be unique under proper assumptions, which are discussed later.

The SU(5) gauge fields at each of the three copies of $M^{4}$ are calculated as follows:
\begin{align}
A(x,1) &= \sum_{i} a_{1}^{i \, \dagger} d  a_{1}^{i} =  i T^{a} A_{1}^{a} (x) \equiv A,  \\
A(x,2) &= \sum_{i} a_{2}^{i \, \dagger} d  a_{2}^{i} = i T^{a} A_{2}^{a} (x) \equiv P A P, \label{A2}\\
A(x,3) &= \sum_{i} a_{3}^{i \, \dagger} d  a_{3}^{i} = 0,
\end{align}
where $T^{a} (a=1, \cdots, 24)$ are the generators of SU(5), 
provided that the following traceless condition is imposed:
\begin{equation}
{\rm Tr} A(x,1) = {\rm Tr} A(x,2) = 0.
\end{equation}

In order to determine Higgs fields, we set the matrix $M_{nm}$ as 
\begin{align}
M_{12} &= M_{21} = M {\rm diag} (1,1,1,1,1) \equiv \Sigma_{0} , \\
M_{13} &= M_{23} = M_{31}^{\dagger} = M_{32}^{\dagger} = 
\mu \begin{pmatrix} 0 & 0 & 0 & 0 & 1 \end{pmatrix}^{T} \equiv H_{0},
\end{align}
where $M (\mu)$ is the energy scale at the stage of GUT (SM) symmetry breaking SU(5) $\to$ SM (SM  $\to$ SU(3)$_{c}$ $\times$ U(1)$_{\rm em}$). 
Thus, the back-shifted Higgs fields are found to be
\begin{align}
\Sigma (x) + \Sigma_{0} &= H_{12} (x) = \sum_{i} a_{1}^{i \, \dagger} M P a^{i}_{1} P = P H_{21} (x) P,  \label{H12} \\
H(x) + H_{0} &= H_{13} (x) = \sum_{i} a_{1}^{i \, \dagger} M_{13} a^{i}_{3} =  P H_{23} (x) . \label{H13}
\end{align}
From Eqs.~(\ref{H12}) and (\ref{H13}), it is found that $H_{13}$ is a 5 $\times$ 1 matrix-valued field transforming like 5 representation under SU(5), and $H_{12}$ is a 5 $\times$ 5 matrix-valued field that is a linear combination of the field like 1 and 24 representation. 
The discussion on the field $\Sigma (x)$ is presented later.
Note that both the parity assignments of the gauge boson (\ref{A2}) and the Higgs boson (\ref{H13}) are determined by the condition of the ``constituent fields'' (\ref{PP}). They are independent conditions in the original orbifold GUT model.

\vspace{18pt}

In order to calculate the Lagrangian (\ref{finalL}), 
it is necessary to consider which $X'_{nml}(x)$ are independent of the Higgs field and which $X'_{nml}(x)$ are not. Here we refer to only dependent $X'_{nml}(x)$, which should be kept in the Lagrangian:
\begin{align}
X_{121}' &= X_{212}' = \sum_{i} a_{1}^{i \, \dagger} M_{12} M_{21} a^{i}_{1} = M^{2}, \\
X_{313}' &=X_{323}' = \sum_{i} a_{3}^{i \, \dagger} M_{31} M_{13} a^{i}_{3} = \mu^{2}, \\
X_{123}'  & = PX_{213}' = \sum_{i}  a_{1}^{i \, \dagger} M_{12} M_{13} a^{i}_{3} = M (H_{1}+H_{0}), \\
X_{312}'  & = X_{321}' P = \sum_{i} a_{3}^{i \, \dagger} M_{31} M_{12} a^{i}_{2} = M (H_{1}+H_{0})^{\dagger} P.
\end{align}
Since other $X_{nml}'(x)$ are auxiliary fields independent of the Higgs, 
the corresponding $P_{nm}(x)$ and $Q_{nml}(x)$ are entirely eliminated in Eq.~(\ref{finalL}) by  $\partial \L / \partial X_{nml}' = 0$.  
Substituting these results into Eq.~(\ref{finalL}), we find the final form of the Lagrangian: 
\begin{align}
\L_{\rm SU(5)} = &- {1\over 4 g^{2} } \tr F_{\mu\nu}^{\dagger} F^{\mu\nu} 
  +{\alpha ^{2} \over 2 g^{2}} \tr | D_{\mu} (\Sigma + \Sigma_{0}) |^{2} 
+ \bigg ( {\alpha^{2} \over 2 g^{'2} } + {\beta^{2} \over 2 g^{2}} \bigg )  | D_{\mu} (H + H_{0}) |^{2}  \nonumber \\
&-{\alpha^{4} \over 2 g^{2}} \tr ||\Sigma + \Sigma_{0}|^{2} - M^{2} |^{2} 
-{\alpha^{2} \beta^{2} \over 2 g^{'2}} ||H + H_{0}|^{2} - \mu^{2} |^{2} \nonumber \\ 
&- \lsp { \alpha^{2} \beta^{2} \over 2 g^{2}} + {\alpha^{4} \over 2 g^{'2} } \rsp | (\Sigma P + MP - M) (H + H_{0})|^{2}. \label{SU5}
\end{align}
Here, $F_{\mu\nu} = \partial_{\mu} A_{\nu} - \partial_{\nu} A_{\mu} + [A_{\mu} , A_{\nu}]$,
and we have assumed the $1 \getsto 2$ symmetry $g_{1}^{2} = g_{2}^{2} = g^{2}$, 
$\alpha_{1}^{2} = \alpha_{2}^{2} = \alpha^{2}$, and set $g_{3}^{2} = g^{'2}, \alpha_{3}^{2} = \beta^{2}$. 

In particular, the mass term of the 5 representation Higgs is computed as
\begin{equation}
| (MP - M) H |^{2} = M^{2} {\rm diag}(4,4,4,0,0) \, H^{\dagger} H.
\label{higgsmass}
\end{equation}
%,
Similarly, the gauge boson masses are
\begin{align}
| D_{\mu} H_{12} |^{2} &\ni  (A_{\mu} M - M P A_{\mu} P)^{2} = (M A_{\mu}^{\hat a} T^{\hat a})^{2} ,
\label{gaugemass}
\end{align}
where $\hat a$ runs the broken generators except those of SU(3)$_{c}$ $\times$ SU(2)$_{L}$ $\times$  U(1)$_{Y}$.
Equations~(\ref{higgsmass}) and (\ref{gaugemass}) show that the parity assignment condition (\ref{PP}),  $a^{i}_{2} = P a^{i}_{1} P$, invokes SU(5) symmetry breaking, and provides the colored triplet Higgs and broken gauge bosons with heavy mass of order $M$.
Therefore, it is reasonable to consider that the symmetry breaking by the condition (\ref{PP}) corresponds to the orbifold GUT mechanism \cite{Kawamura:2000ev} in the NHM. This is our main result. Of course, this result holds only at tree-level. However, if the parity symmetry imposed between the first and second sheets is not broken by the quantum correction, mixing between the Higgs doublet and triplet is prohibited at the quantum level.

%%%%%%%%%%%%%%%%%%%%
\subsection{Discussions}
%%%%%%%%%%%%%%%%%%%%

By imposing the condition (\ref{PP}), the additional Higgs field $\Sigma(x)$ emerges (in the case of  $P = 1$, $H_{12} (x) = \sum_{i} a^{i}_{1} M a^{i}_{1} = M$ and then $\Sigma(x)$ disappears ). 
To investigate whether it is possible to eliminate the $\Sigma$ field or not, let us consider the most general form of the parity assignment:
\begin{align}
a_{i1} = O a_{i} P, ~~
a_{i2} = Q a_{i} R.
\end{align}
Here we assume that  $O, P, Q, R$ are all diagonal and commutative, 
and $O^{2} = P^{2} = Q^{2} = R^{2}  = 1$, which fulfills the Hermitian condition (\ref{Hermitian}).
$OPQR = 1$ is also imposed so as to give $\Sigma(x) \to 0$ at the proper gauge transformation.
In this case,  gauge symmetry breaking in the Lagrangian (\ref{SU5}) requires the conditions $O \neq Q$ and $P \neq R$. Since only the differences between them are important, we can set $O=P=1$. Then $Q = R$ holds by $OPQR = 1$, which leads to the condition (\ref{PP}). 
On the other hand, in order to eliminate the $\Sigma$ field, $O = Q$ should hold.
Then, these two conditions, gauge symmetry breaking and elimination of the $\Sigma$ field, are incompatible. 
Under these proper assumptions, we conclude that it is impossible to avoid this kind of field to implement the orbifold GUT mechanism in this formalism.

In order to probe the gauge transformation property of this $\Sigma$ field, 
the parity matrix $P$ is decomposed into a linear combination of the hypercharge $Y$ and identity matrix $I$ as follows:
\begin{align}
Y &= {1\over \sqrt{60}} {\rm diag} (2,2,2,-3,-3),  \\
P &=  {\rm diag} (-1,-1,-1,1,1) = - {2\over 5} \sqrt{60} Y - {1\over 5} I \equiv cY + dI .
\end{align}
Then, the Higgs field is found to be
\begin{align}
H_{12} P &= \sum_{i} a_{1}^{i \, \dagger} M P a^{i}_{1} = M \sum_{i} a_{1}^{i \, \dagger} (cY + dI) a^{i}_{1} \\
 &= cM \sum_{i} a_{1}^{i \, \dagger} Y a^{i}_{1} + dM .  \label{cMY}
\end{align}
Since $Y$ is the generator of SU(5), the first term in Eq.~(\ref{cMY}) behaves as an adjoint Higgs field. 

Finally, we will mention the nucleon stability. 
In this toy SU(5) model, heavy gauge bosons of broken symmetry $X_{\mu}, Y_{\mu}$ and colored triplet Higgs $H^{c}$ induce nucleon decay. However, this is an intrinsic problem in SU(5) and we can extend the lifetime of the nucleon by other mechanisms, e.g., supersymmetrization. 
However, couplings between fermions and $X_{\mu},Y_{\mu}, H^{c}$ bosons can be forbidden by the proper parity assignment of fermions in the normal orbifold GUT \cite{Altarelli:2001qj}. 
There is, thus, a possibility that these couplings could also be prohibited in this formalism.
However, the fermionic sector in the NHM has subtleties and there are several definitions \cite{Morita:1993zv,Castellani:2000xt,Chamseddine:1992kv}. We leave the construction of the fermionic sector for future work.

%%%%%%%%%%%%%%%%%%%%
\section{Conclusions}
%%%%%%%%%%%%%%%%%%%%

In this paper, we have implemented the orbifold GUT mechanism in the noncommutative Higgs model. 
An assignment of $Z_{2}$ parity to the  ``constituent fields'' induces parity assignments of both the gauge and Higgs bosons, because these bosons are treated as some kind of composite field in this formalism.
This is the main difference from the original orbifold GUT model, which treats the assignments of bosons as independent conditions. 
As a result, some of the gauge bosons and the colored triplet Higgs boson receive heavy mass comparable to the GUT scale, and the SU(5) gauge symmetry is broken. 
 No particles appear other than the SM ones in the massless states.
In fact, nucleon decay is a problem in this model. 
However, there is a possibility that couplings between fermions and $X_{\mu},Y_{\mu}, H^{c}$ bosons could be forbidden by the proper parity assignment of fermions.  
We leave the construction of the fermionic sector for future work.

%%%%%%%%%%%%%%%%%%%%
\section*{Acknowledgements}
%%%%%%%%%%%%%%%%%%%%

The author would like to acknowledge R.~Kitano for useful discussions and valuable comments.
This study is supported by JSPS Research Fellowships for Young Scientists, No. 24 $\cdot$ 8357.

%\bibliographystyle{bib/h-physrev50}
%\bibliography{bib/preon,bib/NCG,bib/compositeHiggs,bib/ClassicalConformal}

\end{document}